\documentclass[10pt]{article} 
\usepackage[T1]{fontenc}   
\usepackage[utf8]{inputenc}
\usepackage{mathrsfs}
\usepackage[top=1in, bottom=1in]{geometry}  
\usepackage{setspace}      
\usepackage{nicefrac}     
\usepackage{microtype}      
\usepackage{booktabs} 
\usepackage{bbm} 
\usepackage{float}
\usepackage{hyperref}
\usepackage{amsmath}
\usepackage{amssymb} 
\usepackage{graphicx}
\usepackage{subfigure}
\usepackage{caption}
\usepackage{comment}
\usepackage{natbib}
\usepackage{algorithm}
\usepackage{algpseudocode}
\usepackage{comment}

\usepackage{multirow}

\usepackage{tikz}
\usepackage{amsmath, amssymb}
\usetikzlibrary{arrows.meta, positioning, shapes.geometric, shapes.misc, calc, backgrounds}
\title{InnerGS: Internal Scenes Reconstuction and Segmentation via Factorized 3D Gaussian Splatting}
\author{
Shuxin Liang \\
University of Alberta
\and
Yihan Xiao \\
Sichuan University
\and
Wenlu Tang\thanks{Corresponding Author} \\
University of Alberta
}
\date{}
\begin{document}
\maketitle
\begin{abstract}



3D Gaussian Splatting (3DGS) has recently gained popularity for efficient scene rendering by representing scenes as explicit sets of anisotropic 3D Gaussians. However, most existing work focuses primarily on modeling external surfaces. In this work, we target the reconstruction of internal scenes, which is crucial for applications that require a deep understanding of an object's interior. By directly modeling a continuous volumetric density through the inner 3D Gaussian distribution, our model effectively reconstructs smooth and detailed internal structures from sparse sliced data. Beyond high-fidelity reconstruction, we further demonstrate the framework's potential for downstream tasks such as segmentation. By integrating language features, we extend our approach to enable text-guided segmentation of medical scenes via natural language queries. Our approach eliminates the need for camera poses, is plug-and-play, and is inherently compatible with any data modalities. We provide cuda implementation at: \url{https://github.com/Shuxin-Liang/InnerGS}.
\end{abstract}

\section{Introduction}



Three-dimensional (3D) reconstruction has become a central focus in computer vision and graphics, driven by applications in virtual reality, robotics, and medical imaging. Recent approaches like Neural Radiance Fields (NeRF) have revolutionized 3D scene capture, enabling novel-view synthesis and detailed scene representations from 2D images. NeRF models represent a scene as a continuous volumetric field learned by a neural network, achieving impressive flexibility in view synthesis and geometry reconstruction. However, achieving high visual quality with NeRF often requires heavy multi-layer perceptrons that are costly to train and slow to render \cite{kerbl20233dgaussiansplattingrealtime}. These limitations triggers a search for more efficient 3D representation and rendering techniques. 

One emerging solution is 3D Gaussian Splatting (3DGS), which has recently gained popularity for fast neural rendering. Instead of relying on a deep network to implicitly encode the scene, 3DGS represents the scene with a set of explicit anisotropic 3D Gaussians. Critically, this approach offers a breakthrough in speed: it achieves real-time 1080p novel-view rendering ($\geq 30$ fps) and slashes training times from hours to minutes. 3DGS combines the merits of neural radiance fields in high-quality, continuous reconstruction with the efficiency of point-based rendering and sparse computation, yielding a new state-of-the-art in 3D scene reconstruction and view synthesis.


While 3D Gaussian Splatting (3DGS) has traditionally targeted exterior view synthesis, there is a growing focus on the more complex challenge of modeling internal structures. This task marks a fundamental shift from surface completion to volumetric inference, which is vital for applications requiring a deep understanding of an object's interior. For example, fine-grained internal reconstruction is crucial in medical imaging for diagnostics and surgical planning \citep{wang2025neuralradiancefieldsmedical}. It is also essential for robotics and VR, where systems must understand object composition for realistic manipulation and interaction \citep{qiu2024advancingextendedreality3d, zhu20243dgaussiansplattingrobotics}. To address this challenge, researchers are developing innovative techniques. In medical imaging, physics-based attenuation models, in which the radiance field is modulated by tissue-specific attenuation coefficients, have already yielded notable improvements in 
CT reconstruction quality \citep{zha2024r2gaussianrectifyingradiativegaussian}. On another front, incorporating deep diffusion models and self-augmentation strategies into 3DGS pipelines may help overcome limitations of sparse-view data,  providing additional pseudo-supervisory signals that enhance internal texture synthesis \citep{wu2024fruitninja3dobjectinterior}. 

However, significant challenges remain. Most current methods are tailored to X-ray or CT projections and struggle to generalize to other data modality such as 
MRI, 
fMRI, or other large-scale 3D remote sensing data. In addition, they heavily rely on external views and their reconstructions either miss fine-grained interior details or require complex regularization that blurs subtle structures. At last, many hybrid training approaches rely on deep MLPs or per-view diffusion refinement, leading to sub-real-time inference and adding difficulty for building extension on it.


To address these limitations, we propose a novel method specifically designed for modeling internal scenes, which employs 3D Gaussian density instead of traditional projection-based 2D Gaussian rendering. Specifically, we first compute conditional 2D Gaussian splats at each depth slice to determine the 2D Gaussian center and influence radii in the image plane, and then combine these conditional 2D Gaussian with a marginal 1D Gaussian along the depth axis to formulate the complete 3DGS density. This allows us to perform the same tile-based rasterization process as in 3DGS, enabling the reconstruction of smooth and detailed internal structures from sparse sliced data.

Our contributions are listed as follows:
\begin{itemize}
    \item We introduce Inner Gaussian Splatting, the framework to leverage 3D Gaussian Splatting for direct volumetric inference. This novel approach enables high-fidelity reconstruction of complex internal structures from sparse, pose-free sliced data. 
    \item We propose an efficient, slice-based rendering pipeline centered on Conditional Splatting. This technique dynamically adapts the sampling of Gaussians for each 2D slice, improving computational efficiency and reconstruction accuracy compared to heuristic projection methods. 
    \item We provide a plug-and-play CUDA implementation of our solution. We further demonstrate the method's effectiveness on several medical datasets. The framework reconstructs static scenes like brain and cardiac MRIs , as well as 4D dynamic sequences such as wrist motion and brain fMRI data. 
\end{itemize}


 


\section{Related Works}

\paragraph{Novel View Synthesis from Sparse Views}
3D Gaussian splatting (3DGS) has emerged as a compelling method for novel view synthesis \citep{kerbl20233dgaussiansplattingrealtime}. However, the reconstruction quality of 3DGS strongly depends on the quality of input views, making it vulnerable when views are sparse or unposed. To address these challenges, COLMAP-Free \citep{fu2024colmapfree3dgaussiansplatting} 
Moreover, end-to-end method directly predicting 3D Gaussian parameters from sparse inputs through implicit neural representations, including Gamba \citep{shen2024gambamarrygaussiansplatting}, Splatter Image \citep{szymanowicz2024splatterimageultrafastsingleview}, and pixelSplat \citep{charatan2024pixelsplat3dgaussiansplats}, which combine U-Net backbones, transformer modules, and attention mechanisms to regress Gaussian centers, opacities, and spatial parameters. Diffusion models have been integrated into the 3DGS pipeline to hallucinate novel views or refine geometry under sparse supervision. Frameworks such as Posediffusion \citep{wang2024posediffusionsolvingposeestimation},  DreamGaussian \citep{tang2024dreamgaussiangenerativegaussiansplatting}, Deceptive-NeRF/3DGS \citep{liu2024deceptivenerf3dgsdiffusiongeneratedpseudoobservationshighquality} incorporate generated views from the diffusion model into the training, contributing to improved visual fidelity in sparse settings.


Existing work extend it to sparse settings and single view, however, there's limited work on \textbf{sparse view with sliced data}, which pose fundamentally different challenges requiring volumetric inference rather than surface completion. Our work addresses this gap by reconstructing \textbf{dense internal structures} from slices without relying on multi-view poses. 


\paragraph{Internal Structure Reconstruction with NeRF and 3DGS} 
Internal structure reconstruction aims to recover the hidden, volumetric details of objects from external 2D images.  In medical imaging, MedNeRF \citep{coronafigueroa2022mednerfmedicalneuralradiance} adapts NeRF for reconstructing 3D-aware CT projections from single X-rays, introducing physics-informed priors to capture attenuation properties of tissues. Similarly, NAF \citep{Zha_2022} models 3D attenuation as a continuous neural field and uses a self-supervised discriminator and data augmentation to generate high-quality 3D-aware CT projections from a single X-ray. SAX-NeRF \citep{cai2024structureawaresparseviewxray3d} introduces a Line Segment-based Transformer and specialized ray sampling for sparse-view X-ray reconstruction, improving the capture of internal structural details from limited views. The survey by \citet{wang2025neuralradiancefieldsmedical} comprehensively review NeRF developments and highlight adaptations for volumetric and internal reconstructions.

Gaussian Pancakes \citep{bonilla2024gaussianpancakesgeometricallyregularized3d} introduces geometrically-regularized splats for realistic endoscopic reconstruction, demonstrating 3DGS’s suitability for internal anatomical modeling. Likewise, Multi-Layer Gaussian Splatting \citep{kleinbeck2024multilayergaussiansplattingimmersive} explores immersive anatomy visualization, revealing how layered Gaussian representations can capture internal structures. Recent innovations include 
FruitNinja \citep{wu2024fruitninja3dobjectinterior}, which leverages Gaussian splats to generate internal textures of objects when virtually sliced. Focusing on Medical Application, X-Gaussian \citep{cai2024radiativegaussiansplattingefficient} replaces spherical harmonics with radiation intensity response functions (RIRFs) to model the isotropic nature of X-ray attenuation, achieving faster inference and superior image quality compared to NeRF-based approaches. DDGS \citep{gao2024ddgsctdirectiondisentangledgaussiansplatting} further refine radiative modeling by separating isotropic attenuation from minor anisotropic effects such as scattering and beam hardening. With structural prior, X-GRM \citep{liu2025xgrmlargegaussianreconstruction} fixes Gaussian centers on a voxel grid while learning radiative attributes via transformer networks, achieving stable convergence and superior sparse-view CT reconstructions.

However, existing work focus on X-ray and CT data, hard to generalize and \textbf{adapt to multiple modality}, including but not limited to MRI, fMRI, LiDAR and point cloud. In contrast, our work is \textbf{plug-in-and-play} with considerable potential of integration with existing 3DGS application. 


\section{Method}
\begin{figure}[htbp]
    \centering
    \includegraphics[width=\textwidth]{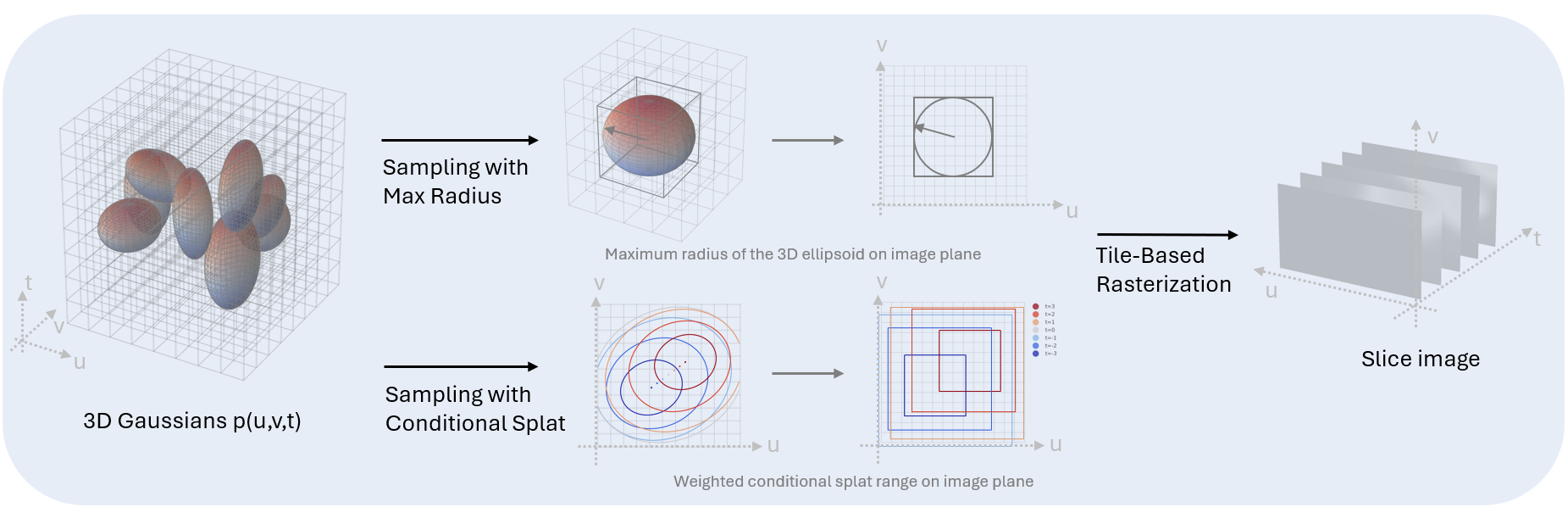}
    \caption{ Illustration of the two sampling method for rasterization: (Top) 3D Ellipsoid Projection computes a cube and project identical bounding boxes across slices, while (Bottom) Conditional Splatting adapts the bounding box per slice based on conditional Gaussians.}
    \label{fig:pipeline}
\end{figure}

Our method introduces a photorealistic scene representation tailored for scenes rich in interior detail. This section proceeds as follows. Section \ref{internal} introduces our Inner 3D Gaussian Splatting framework, which models volumetric density for reconstructing internal structures from sparse slices. Section \ref{optimization} describes the rendering and optimization process. An overview of the pipeline is shown in Figure \ref{fig:pipeline}.

\subsection{Preliminary: 3D Gaussian Splatting}\label{preliminary}

In 3D Gaussian Splatting, a scene is represented as a cloud of 3D Gaussians. Each Gaussian has a theoretically infinite scope and its influence on a given spatial position $\mathbf{x} \in \mathbb{R}^3$ defined by an unnormalized Gaussian function:
\begin{equation}
p(\mathbf{x}) = e^{-\frac{1}{2} (\mathbf{x} - \mu)^T \Sigma^{-1} (\mathbf{x} - \mu)},
\end{equation}

Within this setup, the position of each Gaussian is denoted by the vector $\mu = (\mu_x, \mu_y, \mu_z)$. Its covariance matrix $\Sigma$ is decomposed into a scaling component and a rotation component, formulated as $\Sigma = RSS^\top R^\top$, where $S$ is a diagonal matrix containing scale factors $s_x$, $s_y$, and $s_z$, and $R$ is derived from a unit quaternion $q$ that encodes rotation. Additionally, each 3D Gaussian incorporates coefficients of spherical harmonics (SH) to model view-dependent color variations, as well as an opacity parameter $\alpha$. These parameters are jointly optimized by minimizing a rendering loss during training.

\subsection{3D Gaussian for Internal Scenes}\label{internal}
\paragraph{Problem formulation}


In the standard 3DGS framework, each Gaussian is projected onto the 2D image plane. This projection yields a 2D elliptical splat, whose mean and covariance are derived from the 3D Gaussian transformed by the camera’s view and projection matrices. Let $(u,v)$ denote pixel coordinates on the image plane, where $u \in [0, W-1]$ and $v \in [0, H-1]$ for an image of width $W$ and height $H$. The screen-space contribution of the $i$-th splat at pixel $(u,v)$ is computed as a density $p_i(u,v)$, multiplied by its opacity $\alpha_i$ and color $c_i$. Gaussians are rendered in front-to-back order that accumulates color and opacity to produce the final pixel color, as described in Eq.~(\ref{eq:render}), where $N$ denotes the total number of Gaussians.

\begin{equation}
\label{eq:render}
I(u,v) = \sum_{i=1}^N p_i(u,v) \alpha_i c_i \prod_{j=1}^{i-1} \left( (1 - p_j(u,v)) \alpha_j \right).
\end{equation}

However, this projection-based mechanism is fundamentally unsuitable for modeling internal volumetric structures for two primary reasons. First, the learned Gaussians are concentrated on the object's surface, leaving the interior volume largely empty or undefined. Second, the projection-based rendering pipeline relies on camera parameters (view and projection matrices) which are absent in sliced data. Consequently, applying the original 3DGS to sparse slices fails to capture the continuous volumetric information between the slices, leading to incomplete internal representations. 

\paragraph{Representation of Inner 3D Gaussian}

To overcome these limitations, our goal is to define a scene representation that can directly model volumetric density. We consider any pixel $(u,v)$ on a 2D slice,  where the slice is located at a depth $t$ along a given axis. Instead of relying on projection, we calculate the contribution of each Gaussian to pixel as $p_i(u, v, t)$, thus, the final color of the pixel is computed as: 

\begin{equation}
\label{eq:conditional_render}
I(u, v, t) = \sum_{i=1}^N p_i(u, v, t) \alpha_i c_i \prod_{j=1}^{i-1} \left( 1 -  p_j(u,v,t) \alpha_j \right).
\end{equation}

Through this formulation, we directly model a continuous volumetric density, enabling the synthesis of novel slices at arbitrary depths. This inner 3D representation benefits from both continuous (NeRF) and discrete modeling (Mesh): The continuous nature of Gaussians allows seamless volumetric interpolation, enabling the reconstruction of smooth and detailed internal structures. Meanwhile, representing the volume as a discrete set of Gaussians provides an explicit and explainable representation, making the method efficient and suitable for applications such as medical imaging.

\subsection{Rendering and Optimization}\label{optimization}

During rendering, 3DGS avoids computing all Gaussians at every pixel. Instead, each Gaussian is projected onto the image plane as a 2D elliptical splat, and its 3$\sigma$ extent is computed to define a bounding box that identifies overlapping pixels. Consequently, only a subset of candidate Gaussians is sampled for each pixel, significantly reducing computational cost. To enable efficient candidate sampling here, we propose two methods: 3D Ellipsoid Projection and Conditional Splatting.

\paragraph{Method 1: 3D Ellipsoid Projection}
In this method, we approximate the range of a 3D Gaussian as a sphere, whose radius is determined by the maximum axis length of the Gaussian ellipsoid. For a Gaussian with covariance 
\(\boldsymbol{\Sigma}\), 
the ellipsoid’s principal axes align with the eigenvectors of 
\(\boldsymbol{\Sigma}\), 
and their lengths are given by the square roots of the eigenvalues, scaled by a constant:
$$r_{\text{max}} = 3 \cdot \sqrt{\max(\lambda_1, \lambda_2, \lambda_3)},$$
where \(\lambda_1, \lambda_2, \lambda_3\) are the eigenvalues of \(\boldsymbol{\Sigma}\). A cube of edge length \(2\,r_{\text{max}}\)  is centered at the Gaussian’s mean and projected orthogonally onto each image slice. This yields identical 2D bounding boxes across all slices, as shown in Fig.~\ref{fig:pipeline}, which can result in overly large regions being processed and redundant computation.

\paragraph{Method 2: Conditional Splatting} In the second approach, we adopt a conditional formulation to compute the range of each Gaussian on individual slices, demonstrated on the bottom of Fig \ref{fig:pipeline}. Specifically, we splat a point $p_i(u,v,t)$ on Gaussian onto the 2D plane given depth $t$ by a factorized Gaussian structure:
$$p_i(u,v,t)=p_i(u,v|t)p_i(t), $$
where  $p_i(u,v|t)$ is a conditional two-dimensional Gaussian describing the lateral spatial distribution on the image plane, given the slice location $t$, and  $p_i(t)$ is a one-dimensional Gaussian modeling the uncertainty or distribution along the depth axis. As shown in Appendix \ref{parameter}, the conditional mean $\mu_{u,v|t}$ shifts with the slice depth $t$, reflecting the 2D splat center moves across slices. Moreover, by introducing a distance-dependent scaling factor, the extent of the 2D splat decreases with increasing distance between the 3D Gaussian center and the slice. Unlike the first method, which projects the same cubic bounding box onto every slice, this approach computes an adapted bounding box for each slice, leading to more efficient sampling.

With a set of candidate Gaussians sampled for each pixel, we sort them according to the distance between their centers and the image plane. With the differential rasterization process, the subsequent optimization pipeline remains identical to the original 3DGS, leading to the complete training algorithm described in Alg. \ref{alg:conditional_3dgs_training}

\section{Experiment}
\subsection{Simulation Analysis}
We conducted a simulation to compare the two Gaussian selection methods, for identifying candidate Gaussians in a 3D volume of size $20 \times 20 \times 20$. We randomly generated $N=50$ Gaussians with means uniformly distributed between $[5,15]$ in each axis. A density threshold of $0.01$ is used to decide whether a Gaussian should be considered active at a given pixel. 

The metrics collected include: the average bounding-box area (Avg bbox area) per Gaussian; false positives per pixel (FP/pixel), representing the average number of Gaussians with low density that are incorrectly included in the pixel’s candidate set; false negatives per pixel (FN/pixel), representing the average number of Gaussians with high density that are missed by the pixel’s candidate set; the average number of candidate Gaussians per pixel (Cand/pixel); and the rendering time for each.

Table \ref{tab:simulation} shows that Method 2 significantly reduces the average bounding box size (63.09 vs. 209.80) and false positives per pixel (1.82 vs. 15.78), indicating much tighter and more efficient bounding boxes. However, Method 2 introduces some false negatives (0.126 per pixel on average), which Method 1 completely avoids. Method 2 also processes fewer candidates per pixel (2.62 vs. 16.70) and is faster overall (0.10s vs. 0.69s). Based on the result, we adopt Method 2 for real-data experiments.

\begin{table}[H]
\centering
\begin{tabular}{lcccccccc}
\toprule
\textbf{Method} & \textbf{Avg BBox Area} & \textbf{FP/pixel} & \textbf{FN/pixel} & \textbf{Cand/pixel} &  \textbf{Render Time (s)} \\
\midrule
\textbf{M1} & 209.8038 & 15.7871 & 0.0000 & 16.6999 & 0.6859 \\
\textbf{M2} &  63.0913 &  1.8237 & 0.1259 &  2.6166 & 0.1029 \\
\bottomrule
\end{tabular}
\caption{Comparison between Method 1 and Method 2 on bounding box area, error metrics, candidate counts, and timing.}
\label{tab:simulation}
\end{table}

\subsection{MRI Reconstrution}


We evaluate our Inner 3DGS reconstruction method on real-world MRI data in NIfTI format. Specifically, we extract 2D slices along the axial, sagittal, and coronal planes, convert them to RGB format, and normalize the pixel values to the $[0, 1]$ range. To ensure fair spatial coverage, we uniformly sample $5\%$ of slices from each plane for testing, while the remaining slices for training.

All experiments are conducted on a single NVIDIA RTX 4070 GPU. Unless otherwise specified, we use a uniform 3D grid of $42^3$ points as the initial positions for the Gaussian. The model is trained for a maximum of $1{,}000$ iterations or until the absolute loss decrease over 100 iterations falls below $1 \times 10^{-4}$. We report the time required to reach convergence and evaluate the reconstruction quality using Peak Signal-to-Noise Ratio (PSNR) and Structural Similarity Index Measure (SSIM).

\paragraph{Brain MRI} 
We used the BrainWeb Simulated Brain Database \citep{cocosco2016brainweb}, specifically the normal brain phantom with T1-weighted contrast. To reduce data size and adjust spatial resolution, we downsampled the volume by a factor of 0.8 along all three axes. The final images have resolutions of $174 \times 145$ pixels (axial), $145 \times 145$ pixels (coronal), and $145 \times 174$ pixels (sagittal). 

With the experimental setup described above, the reconstruction results are summarized in Table~\ref{tab:brain_results}, and qualitative reconstruction results of axial and sagittal views are shown in Figure \ref{fig:brain}. A detailed comparison between the rendered outputs and ground truth is provided in Appendix \ref{fig:brain_all}. Our method successfully reconstructs fine anatomical structures, such as cortical folds and deep brain boundaries, while preserving smooth transitions in homogeneous areas like white matter or cerebrospinal fluid.

\begin{table}[h]
\centering
\begin{tabular}{lcccc}
\hline
 & \textbf{Axial} & \textbf{Coronal} & \textbf{Sagittal} & \textbf{Average} \\ \hline
\textbf{PSNR (dB)} & 32.4796 & 31.8621 & 33.0792 & 32.4736 \\ 
\textbf{SSIM}      & 0.9664  & 0.9618  & 0.9746  & 0.9676 \\
\hline
\end{tabular}
\caption{SSIM and PSNR metrics for each axis and overall average. The model converged after approximately \textbf{25.9} minutes of training.}
\label{tab:brain_results}
\end{table}

\begin{figure}[h]
    \centering
    \includegraphics[width=\textwidth]{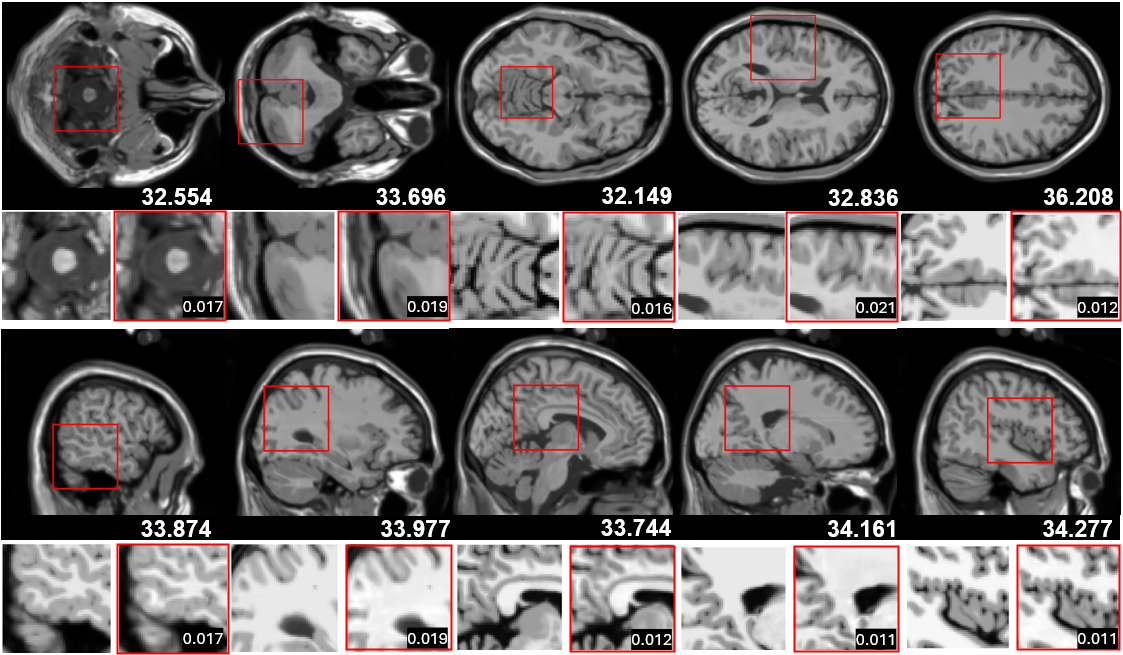}
    \caption{\textbf{Brain MR reconstructions in axial (top) and sagittal (bottom) views.} The top row shows test set renderings with PSNR values overlaid. Red boxes highlight regions for detailed reconstruction. In each zoom-in region, the left patch is the ground truth, and the right is the prediction, with L1 errors indicating absolute differences within the region.}
    \label{fig:brain}
\end{figure}

\paragraph{Cardiac MRI} 
We employ the HVSMR-2.0 dataset \citep{pace2024hvsmr}, Version 2, obtained from Figshare \footnote{(doi: 10.6084/m9.figshare.c.7074755.v2)} and licensed under CC BY 4.0, which provides 3D cardiovascular MR images.
Each volume captures the whole-heart anatomy in a static phase. After extracting 2D slices along the three anatomical planes, the resulting images have typical resolutions of $127 \times 207$ pixels (axial), $127 \times 141$ pixels (coronal), and $207 \times 141$ pixels (sagittal). 

\begin{figure}[H]
    \centering
    \includegraphics[width=\textwidth]{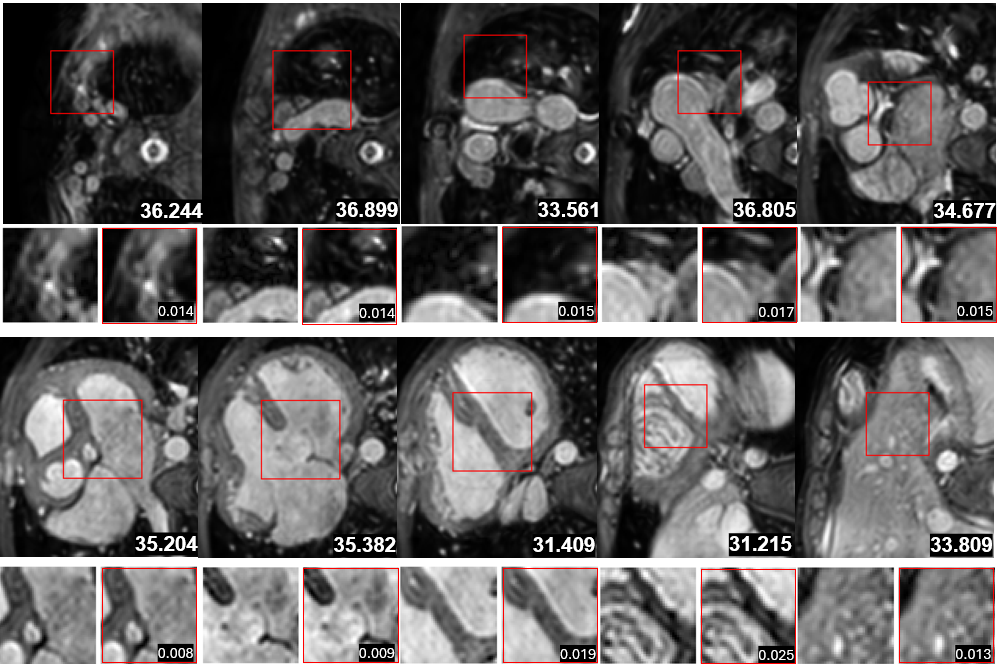}
    \caption{\textbf{Cardiac MR reconstructions (coronal view).}}
    \label{fig:heart_coronal}
\end{figure}

For Cardiac MRI, the quantitative results are summarized in Table \ref{tab:heart_results}. Figure \ref{fig:heart_coronal} presents detailed reconstruction results on the coronal plane, while results for the axial and sagittal plane are reported in Appendix \ref{fig:heart_axial_sagittal}. The  renderings accurately capture details such as the heart chambers and vessels.

\begin{table}[h]
\centering
\begin{tabular}{lcccc}
\hline
 & \textbf{Axial} & \textbf{Coronal} & \textbf{Sagittal} & \textbf{Average} \\ \hline
\textbf{PSNR (dB)} & 30.9107\textsuperscript{*} & 34.2207 & 29.3602\textsuperscript{*} & 31.4972 \\ 
\textbf{SSIM}      & 0.9433  & 0.9539  & 0.9344  & 0.9439 \\
\hline
\end{tabular}
\caption{SSIM and PSNR metrics for each axis and overall average.  The model converged after approximately \textbf{30.4} minutes of training.
\textsuperscript{*}Axial and sagittal views include images affected by brightness shifts from the training set, resulting in slightly lower PSNR values; detail analysis are provided in the Appendix.}
\label{tab:heart_results}
\end{table}

\subsection{Dynamic Reconstruction}
Beyond static anatomical reconstruction, we demonstrate the versatility of our framework by extending it to dynamic Magnetic Resonance Imaging data. This task presents unique challenges: modeling the temporal variations of anatomical structures and physiological activity from time-resolved MRI sequences. Our goal is to faithfully reconstruct these dynamic variations across full 3D volumes from sparsely acquired multi-slice observations.

To address temporal dynamics in diverse dynamic MRI modalities, we adapt the 4D Gaussian Splatting framework \citep{wu20244dgaussiansplattingrealtime}. A static set of 3D Gaussians represents the canonical anatomy, while a learned deformation field network predicts per-frame changes in position, covariance, color, and opacity using a spatiotemporal encoder and a lightweight MLP. We implement a custom data loader to synchronously process all slices at each timestamp. The Gaussians are initialized by pretraining on slices from the first frame, providing a stable anatomical prior. This initialization is then refined by the deformation network to capture temporal variations across frames.



\paragraph{Wrist Motion MRI} 

We validate our method on the Dynamic MRI of the Wrist dataset, Version 2, obtained from Mendeley Data \footnote{doi: 10.17632/9kx5xp7h6d.2} under the CC BY 4.0 license \citep{sharafi2025development}, with 38 temporal frames, each containing 12 wrist slice views at a resolution of 64 × 64.
We use 33 frames for training and hold out every 8th frame for testing. As a strong static baseline, we use the ground-truth image from the preceding timestamp (i.e., test time $t{-}1$) as the prediction for the current test frame $t$. This setup simulates an oracle model that perfectly memorizes and reuses the prior anatomical structure without modeling temporal dynamics. We initialize $35^3$ 3D Gaussians in the canonical space and train our model to convergence in approximately 45 minutes.

\begin{figure}[H]
    \centering
    \includegraphics[width=0.8\textwidth]{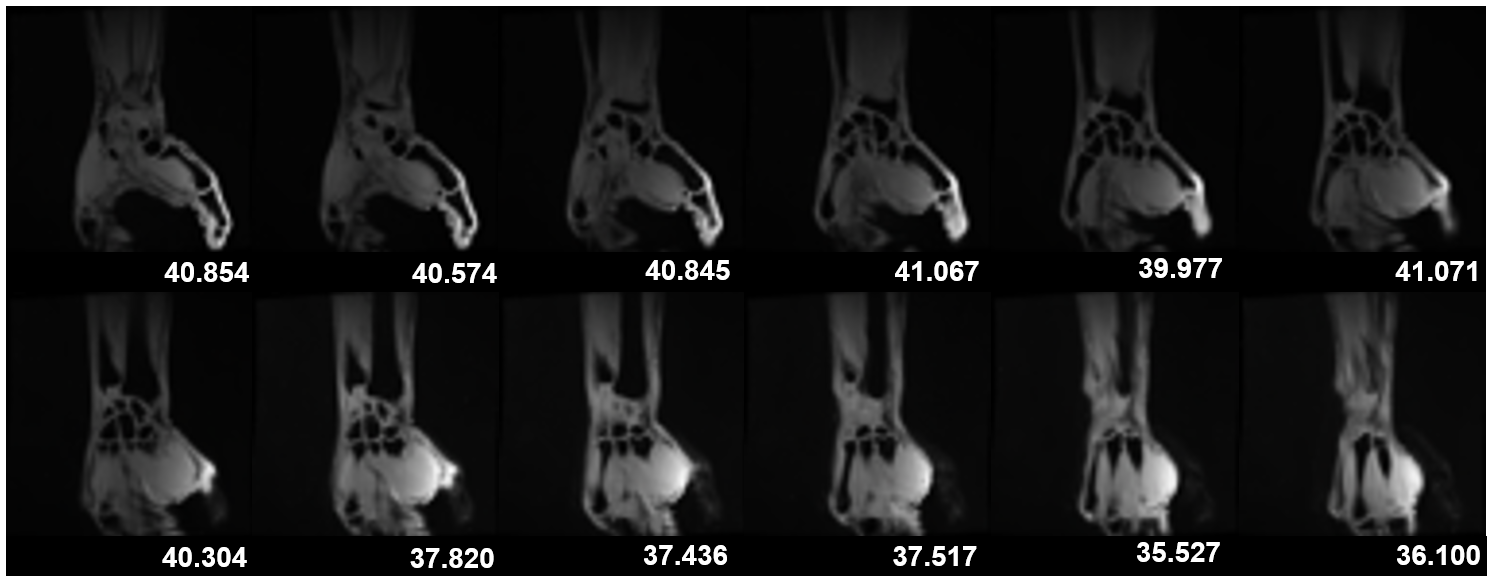}
    \caption{Wrist MR reconstructions at Time = 16 across all axial slices with per-slice PSNR values.}
    \label{fig:hand_axial}
\end{figure}

\begin{figure}[H]
    \centering
    \includegraphics[width=0.8\textwidth]{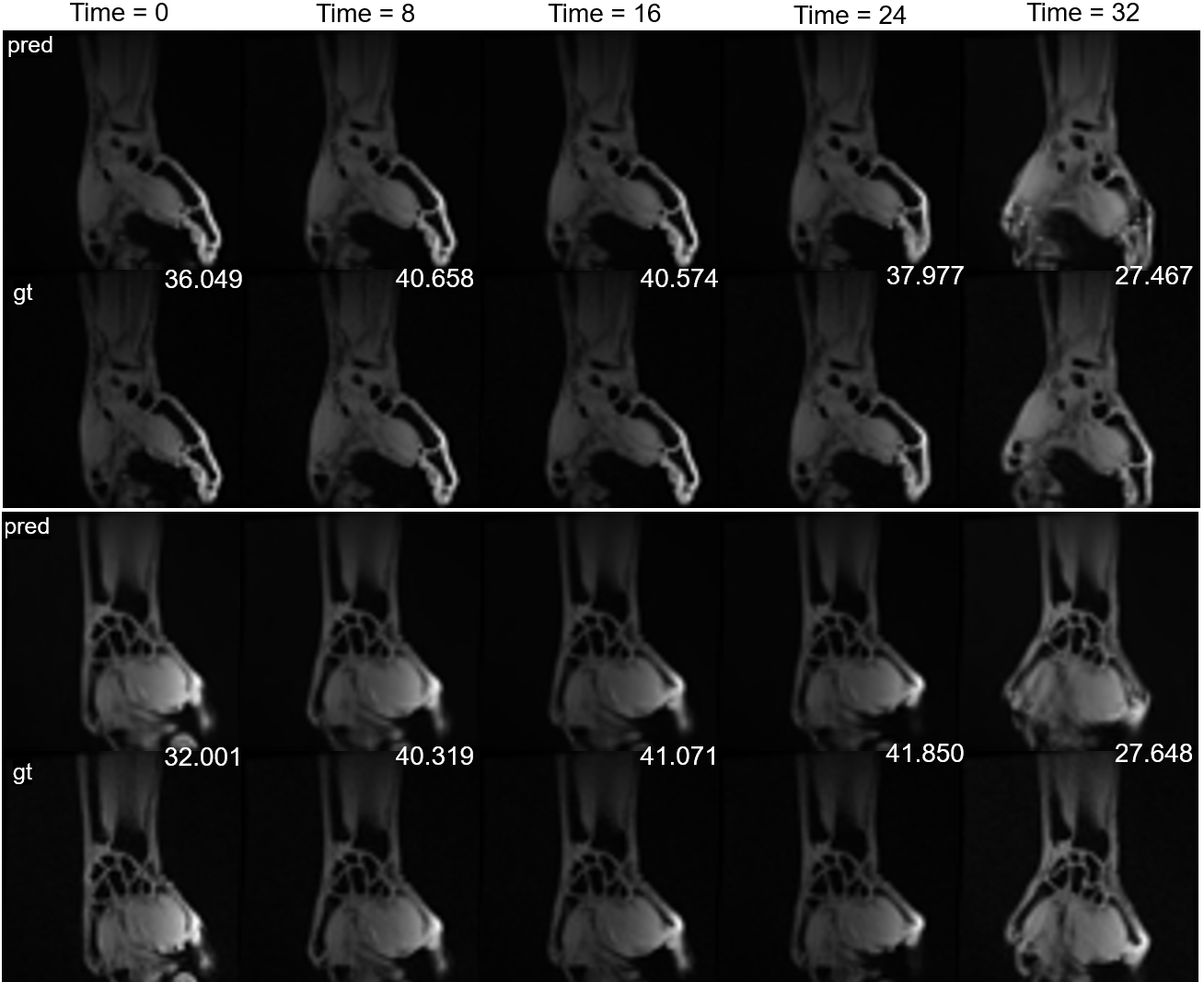}
    \caption{Wrist motion MR reconstructions across test timestamps for two axial slices: \textbf{top} — slice 1, \textbf{bottom} — slice 5. For each slice, predicted reconstructions (pred) are compared with ground-truth images (gt). Our method captures temporally coherent anatomical motion across both shallow and deep wrist structures.}
    \label{fig:hand_temporal}
\end{figure}

Table \ref{tab:wrist_results} presents comparison between our method and this baseline, it shows that our method consistently outperforms the baseline, with particularly large gains at later timestamps with larger anatomical motion. Figure~\ref{fig:hand_axial} shows slice-wise reconstructions at Time~$=16$, illustrating anatomical consistency across the volume. Figure~\ref{fig:hand_temporal} presents predictions over time for two representative slices, highlighting our model’s ability to track wrist motion at different depths. These results highlights our method’s ability to capture nontrivial anatomical changes that static models fail to handle. 

\begin{table}[htbp]
\centering
\begin{tabular}{llcccccc}
\toprule
\textbf{Metrics} & \textbf{Method} & \textbf{Time=0} & \textbf{Time=8} & \textbf{Time=16} & \textbf{Time=24} & \textbf{Time=32} & \textbf{Avg.} \\
\midrule
\multirow{2}{*}{SSIM↑} 
    & Baseline & 0.9466 & 0.9668 & 0.9706 & 0.9396 & 0.8909 & 0.9429 \\
    & \textbf{Ours}     & 0.9443 & 0.9848 & 0.9864 & 0.9865 & 0.9224 & 0.9649 \\
\midrule
\multirow{2}{*}{PSNR↑} 
    & Baseline & 31.4973 & 32.4322 & 34.4965 & 28.6187 & 24.5761 & 30.7242 \\
    & \textbf{Ours}     & 31.0204 & 38.3950 & 39.0909 & 39.0871 & 26.9297 & 34.9046 \\
\bottomrule
\end{tabular}
\caption{Comparison between Baseline (previous frame GT) and Our Method on testing timestamps.}
\label{tab:wrist_results}
\end{table}

\section{Segmentation}
Building upon the potential for downstream clinical tasks, we extended the InnerGS framework to facilitate multimodal medical scene understanding by integrating Vision-Language Models (VLMs). This proposed approach augments each 3D Gaussian with compact language features distilled from a knowledge-enhanced medical VLM \citep{zhao2025largevocabularysegmentationmedicalimages} via a scene-specific autoencoder \citep{qin2024langsplat3dlanguagegaussian}, thereby establishing a direct alignment between visual and textual representations.To ensure computational efficiency, high-dimensional embeddings are distilled into compact latent attributes $f_i$, which are then aggregated using the differentiable splatting equation $F(v) = \sum_{i \in \mathcal{N}} f_i \alpha_i \prod_{j=1}^{i-1} (1 - \alpha_j)$. The rendered latent map is subsequently projected back to the semantic space by a lightweight decoder to enable open-vocabulary alignment. This design significantly circumvents the memory explosion of explicit high-dimensional modeling while preserving volumetric consistency. This mechanism enables precise 3D segmentation of Regions of Interest through open-vocabulary queries. 

Our proposed method demonstrates superior segmentation performance on the CT-ORG subset in SAT-DS \citep{rister2019ctorg, zhao2025largevocabularysegmentationmedicalimages}, achieving a mean Dice Similarity Coefficient (mDSC) of 88.17\%, a mean Intersection-over-Union (mIoU) of 83.72\%, and an overall Pixel Accuracy (PA) of 98.97\%. In specific anatomical regions such as the liver, the model exhibits strong robustness, yielding a DSC consistently above 84.62\%. CT-ORG is licensed under CC BY 3.0.

\begin{table}[htbp]
  \centering
  \caption{Performance of Top 5 Slices}
  \label{tab:top5_slices}
  \begin{tabular}{lccccc} 
    \toprule
    \textbf{Slice ID} & \textbf{PA (\%)} & \textbf{mIoU (\%)} & \textbf{mDSC (\%)} & \textbf{Liver DSC} & \textbf{Others DSC} \\
    \midrule
    Slice 32 & 99.42 & 95.31 & 97.56 & 98.18 & 94.81 \\
    Slice 70 & 98.81 & 94.32 & 97.04 & 97.13 & 94.74 \\
    Slice 29 & 99.20 & 92.72 & 96.14 & 96.26 & 92.60 \\
    Slice 65 & 98.62 & 92.73 & 96.14 & 97.40 & 91.87 \\
    Slice 30 & 99.16 & 92.43 & 96.00 & 94.67 & 93.79 \\
    \bottomrule
  \end{tabular}
\end{table}

\section{Conclusion}
In this work, we introduced a novel extension of 3D Gaussian Splatting tailored for modeling internal
scenes, a task critical for applications such as medical imaging, robotics, and volumetric analysis.
Several exciting avenues for future research remain open. High-fidelity volumetric reconstructions
enabled by our method may facilitate downstream clinical tasks such as detection or segmentation.

\newpage
\appendix
\section{Additional Results}
\begin{figure}[h]
    \centering
    \begin{minipage}[b]{\textwidth}
        \centering
        \includegraphics[width=\textwidth]{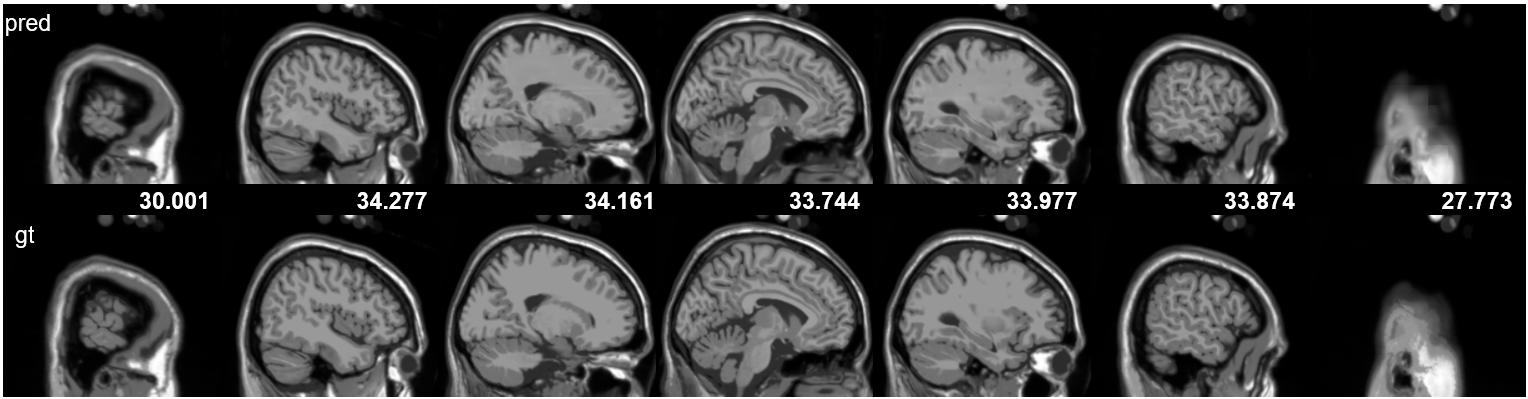}
    \end{minipage}
    
    \vspace{2mm}
    
    \begin{minipage}[b]{\textwidth}
        \centering
        \includegraphics[width=\textwidth]{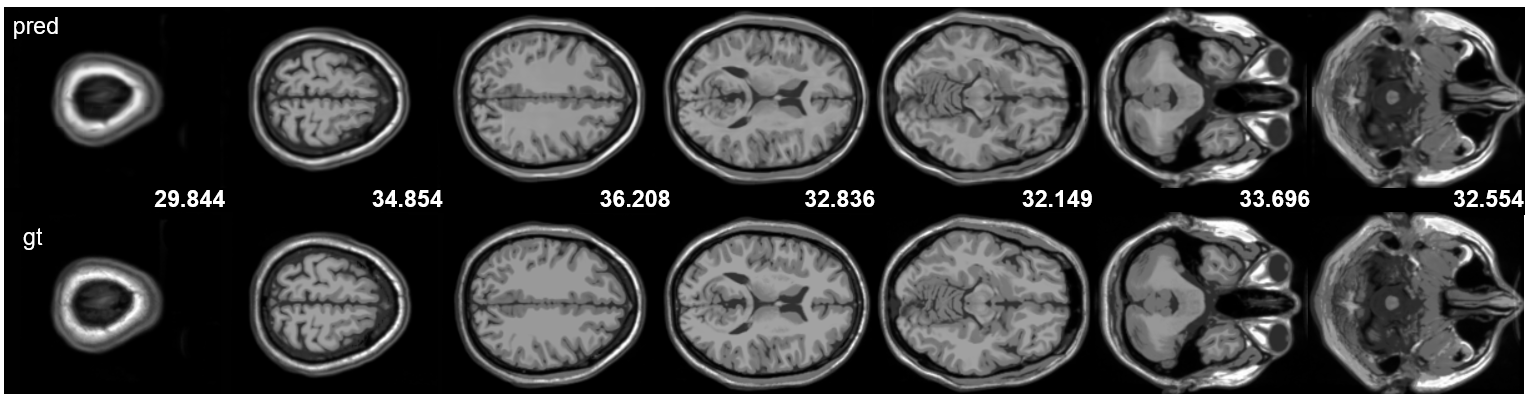}
    \end{minipage}
    
    \vspace{2mm}
    
    \begin{minipage}[b]{\textwidth}
        \centering
        \includegraphics[width=\textwidth]{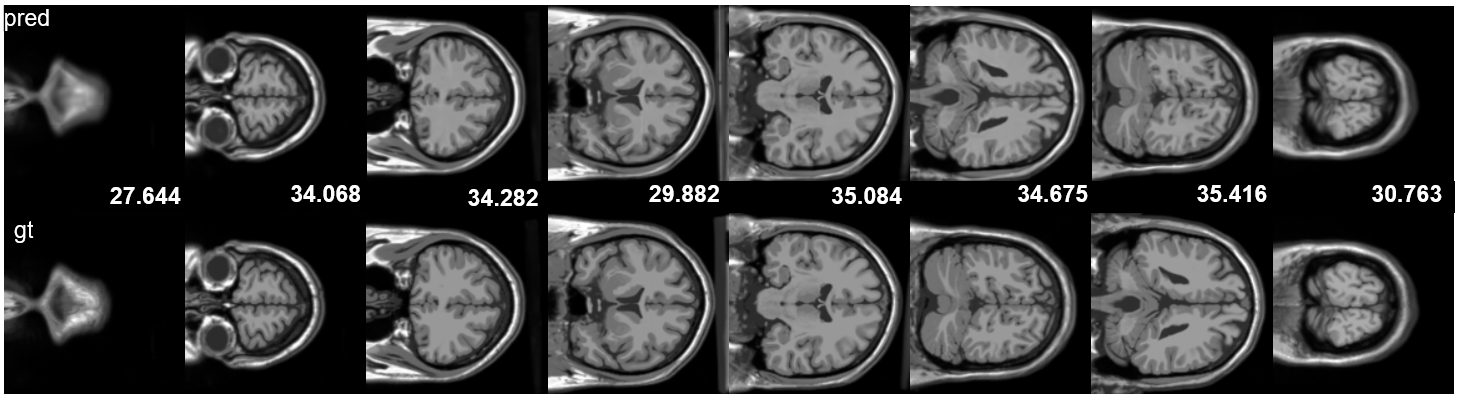}
    \end{minipage}
    
    \caption{\textbf{Brain MR reconstructions in different anatomical planes.} From top to bottom are the axial, sagittal, and coronal views, respectively.}
    \label{fig:brain_all}
\end{figure}

\paragraph{Brain MRI} Figure \ref{fig:brain_all} presents the reconstruction results of brain MRI images in three anatomical planes. From top to bottom, the figure shows the sagittal, axial, and coronal planes, respectively. In each set of images, the top row displays the model’s predicted images (pred), while the bottom row shows the corresponding ground truth images (gt). The predicted images are annotated with their PSNR values to quantify reconstruction quality. It can be observed that the model successfully reconstructs the brain anatomy across all planes, with clear details of gyri and sulci, and demonstrates high consistency with the ground truth in terms of contrast and structural details, indicating strong reconstruction capability.

\begin{figure}[H]
    \centering
    \begin{minipage}[b]{\textwidth}
        \centering
        \includegraphics[width=\textwidth]{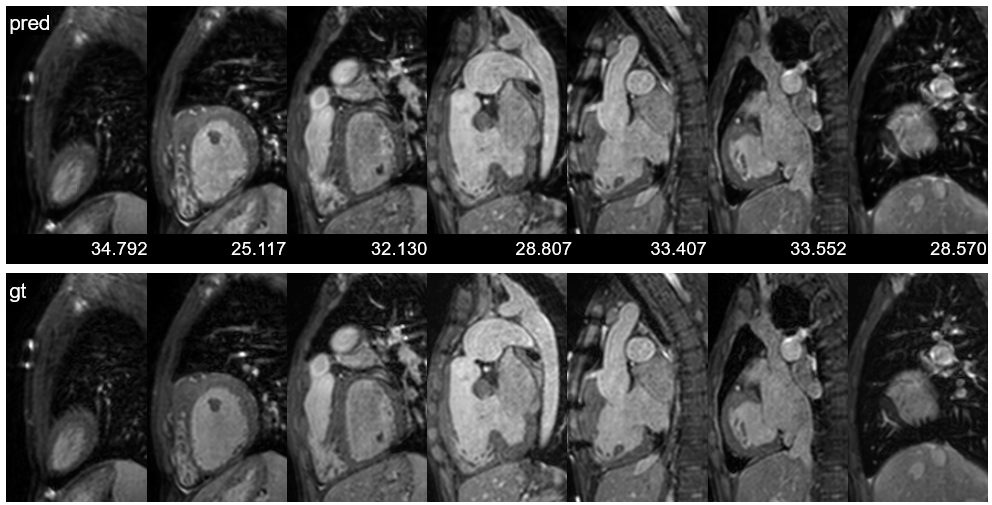}
    \end{minipage}
    
    \vspace{2mm}
    
    \begin{minipage}[b]{\textwidth}
        \centering
        \includegraphics[width=\textwidth]{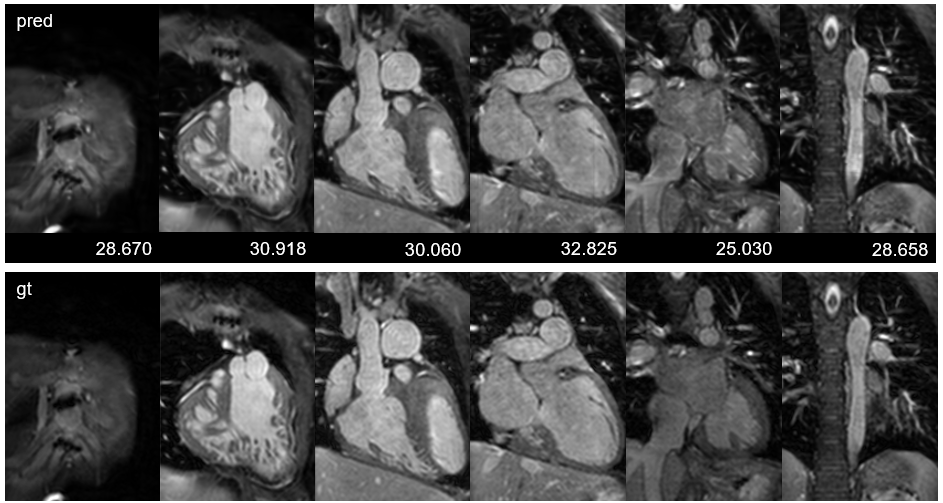}
    \end{minipage}
    
    \caption{\textbf{Cardiac MR reconstructions in axial (top) and sagittal (bottom) planes.} }
    \label{fig:heart_axial_sagittal}
\end{figure}

\newpage
\paragraph{Cardiac MRI} Figure \ref{fig:heart_axial_sagittal} shows the reconstruction results of cardiac MRI images in different anatomical planes (axial and sagittal). It can be observed that the model successfully reconstructs cardiac structures across different planes, accurately depicting anatomical details such as the heart chambers, myocardium, and major vessels. 

\textsuperscript{*}The 2nd axial slice and the 5th sagittal slice showed abnormally low PSNR (25.117 dB and 25.030 dB). Applying affine normalization to match the ground truth’s mean and variance raised PSNR to 35.15 dB and 32.24 dB, respectively. This suggests that the degradation might stems from bias field effects and brightness drift in the training data, rather than structural reconstruction errors.

\newpage
\paragraph{Brain fMRI}

We additionally validate our framework on public functional MRI (fMRI) data from the OpenfMRI project \citep{ds000003:00001}, consisting of 160 temporal frames, each with 33 axial brain slices at a resolution of 64 × 64. This data was obtained from the OpenfMRI database, ds000003.

Each slice is visualized using grayscale intensity mapped from the normalized BOLD signal, which reflects changes in blood oxygenation (BOLD contrast) over time. Unlike dynamic MRI of joints, fMRI captures temporal fluctuations in signal intensity rather than anatomical motion. To adapt our method accordingly, we freeze all geometry-related parameters (e.g., Gaussian positions and covariances) and allow only color and opacity to vary over time. This enables our model to directly capture voxel-wise radiance changes induced by neural activity.

Our training follows a two-stage strategy: we first build a static anatomical structure over 1000 warm-up iterations using the initial frames, followed by 3000 iterations where only color and opacity evolve temporally. We use 140 frames for training and reserve every 8th frame (20 in total) for testing. A custom data loader ensures all slices from a given timestamp are processed synchronously to preserve inter-slice consistency.

\begin{figure}[H]
\centering
\includegraphics[width=\textwidth]{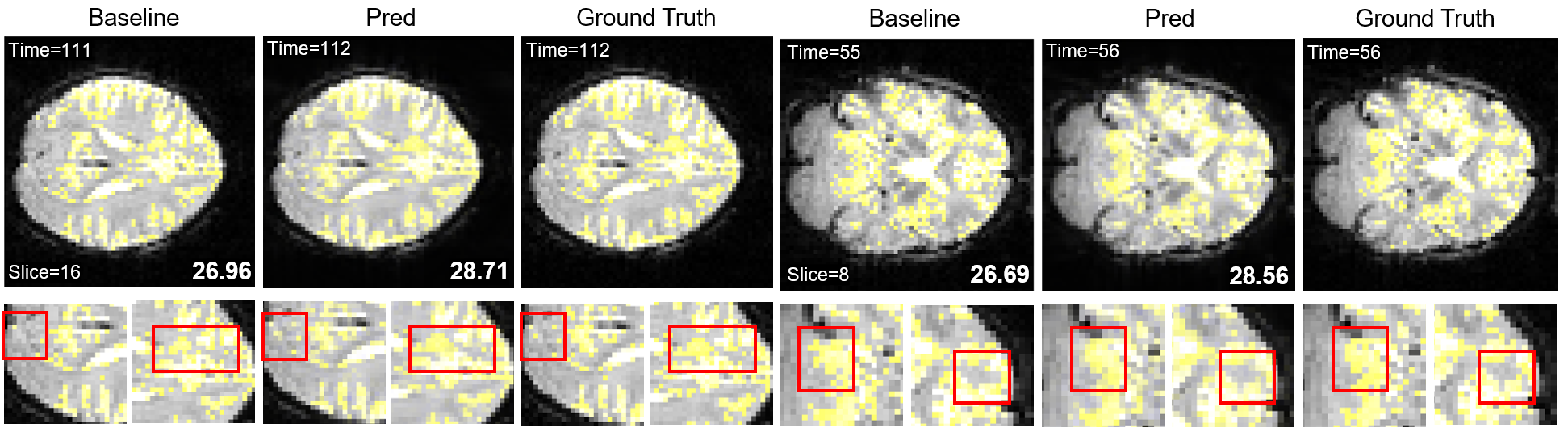}
\caption{\textbf{Qualitative comparison on fMRI slices.} For two representative timepoints and slices, we compare the static baseline (previous frame), our model prediction, and ground truth. Red boxes highlight regions with noticeable signal change. Our method produces sharper and more consistent activation patterns, with higher PSNR (bottom-right).}
\label{fig:fmri}
\end{figure}

\begin{figure}[H]
\centering
\includegraphics[width=\textwidth]{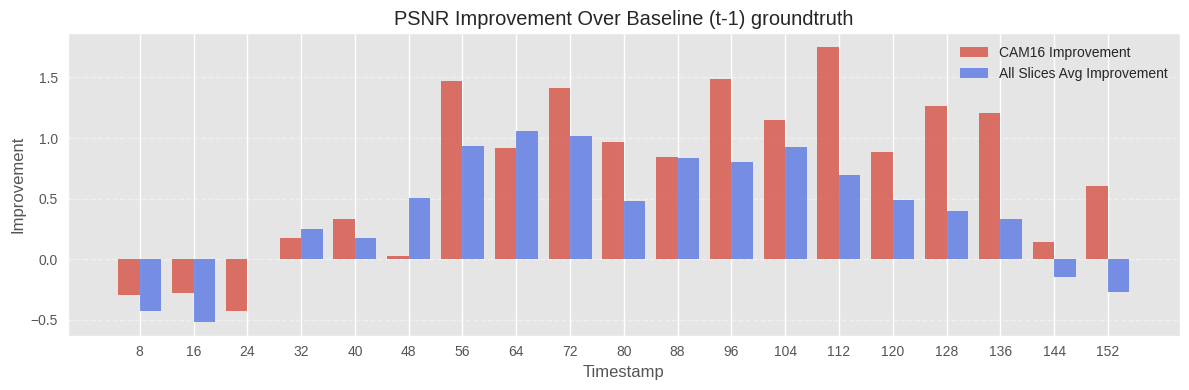}
\caption{\textbf{Improved PSNR across time.} Quantitative improvements in PSNR over the static baseline, plotted over all 20 test frames. We report both the central slice (slice 16), which corresponds to the most active central brain region, and the average over all slices. Our model consistently outperforms the baseline, with the most notable gains observed in this central region where BOLD signal changes are strongest.}
\label{fig:fmri_psnr}
\end{figure}

\newpage
\section{Training Implementation}
Algorithm \ref{alg:conditional_3dgs_training} describes the inner training loop for 3D Gaussian Splatting. It starts by initializing Gaussians on a 3D grid and then iteratively optimizes their parameters using gradient descent. In each step, the algorithm projects Gaussians onto different 2D slice views, computes the rendering and loss, and performs backpropagation. Additional steps like pruning, densification, and cloning are applied to adapt the set of Gaussians dynamically for better scene reconstruction quality.

\begin{algorithm}
\caption{Inner 3D Gaussian Splatting Training}
\label{alg:conditional_3dgs_training}
\begin{algorithmic}[1]
\Require Training slices $\mathcal{D} = \{(I^{\text{axis}}_i, \pi^{\text{axis}}_i)\},\; \text{axis}\in\{x, y, z\}$, $I$ is 2D slice image, $\pi$ is slice position
\Require Grid resolution $g_{\text{res}}$, max iterations $T_{\max}$
\Require Thresholds of opacity, gradient, and scale $(\tau_\alpha,\tau_p,\tau_s)$
\Ensure Optimized Gaussians $\mathcal{G} = \{(\mu,\Sigma,c,\alpha)\}$

\State \textbf{Init:} Build regular grid $\mathcal{P}$ of size $g_{\text{res}}^3$ or SfM estimated point, each Gaussian with $\mu$ around grid points, and default $(s,c,\alpha,q)$
\State Optimizer $\gets \text{Adam}(\theta,\eta)$

\For{$\text{step} = 1$ \textbf{to} $\text{step}_{\max}$}
    \State $L_{\text{total}}\gets0$; Zero gradients: $\nabla\theta\gets0$
    \For{\textbf{axis} $\in\{x,y,z\}$}
        \For{\textbf{each} slice $(I^{\text{axis}}_i,\pi^{\text{axis}}_i)\in\mathcal{D}^{\text{axis}}$}
            \State $\Sigma_n \gets R_n S_n S_n R_n^\top$ \hfill\(\triangleright\) Compute 3D covariances
            
            \State $I_{\text{pred}} \leftarrow \text{Rasterize}(\mu_n, \Sigma_n, \alpha_n, c_n)$
            \Comment{Rendering}
            
            \State $L \gets \text{Loss}(I_{\text{pred}}, I^{\text{axis}}_i)$ \hfill\(\triangleright\) Compute loss
            \State $L_{\text{total}}\gets L_{\text{total}} + L$
        \EndFor
    \EndFor
    \State $L_{\text{total}}.\text{backward()}$ \hfill\(\triangleright\) Back-propagation
    \State $\theta\gets\text{AdamStep}(\theta,\nabla\theta)$

    \If{\text{IsRefinementIteration}$(t)$} 
        \ForAll{Gaussians $(\mu,\Sigma,c,\alpha)\in\mathcal{G}$}
            \If{$\alpha < \tau_\alpha$ \textbf{or} \text{IsTooLarge}$(\mu,\Sigma)$} \hfill\(\triangleright\) Pruning
                \State \text{RemoveGaussian}()
            \ElsIf{$\lVert\nabla_{\mu}L\rVert_2 > \tau_p$} \hfill\(\triangleright\) Densification
                \If{$\lVert\Sigma\rVert_F > \tau_s$} \hfill\(\triangleright\) Over-reconstruction
                    \State \text{SplitGaussian}$(\mu,\Sigma,c,\alpha)$
                \Else \hfill\(\triangleright\) Under-reconstruction
                    \State \text{CloneGaussian}$(\mu,\Sigma,c,\alpha)$
                \EndIf
            \EndIf
        \EndFor
    \EndIf
\EndFor
\end{algorithmic}
\end{algorithm}

\newpage
\section{Parameter of 2DGS and 1DGS}\label{parameter}

Given a 3D Gaussian with mean vector  $\mu^{3D}=[\mu_x, \mu_y, \mu_z]^\top$ and a covariance matrix $\Sigma^{3D} = [(\sigma_{xx}, \sigma_{xy}, \sigma_{xz}), (\sigma_{xy}, \sigma_{yy}, \sigma_{yz}), (\sigma_{xz}, \sigma_{yz}, \sigma_{zz})]
$, we can decompose its probability density into a factorized form along the depth axis~$t$. Specifically, we factorize the joint density $p(u, v, t)$ into the product of a marginal distribution along the depth direction $t$, and a conditional distribution over the 2D coordinates $(u,v)$ given $t$.

The marginal density along the depth axis is given by:
\[
p(t) \sim \mathcal{N}(\mu_z, \sigma_{zz}),
\]
where $\mu_z$ and $\sigma_{zz}$ denote the mean and variance of the Gaussian along the $z$-axis, respectively.

Conditioned on a specific depth $t$, the distribution over the in-plane coordinates $(u,v)$ is Gaussian:
\[
p(u,v \mid t) \sim \mathcal{N} \bigl( \mu_{uv \mid t}, \; \Sigma_{uv \mid t} \bigr),
\]
where the conditional mean $\mu_{uv \mid t}$ is shifted linearly according to the offset $t - \mu_z$:
\[
\mu_{uv \mid t}
=
\begin{bmatrix}
\mu_x \\ \mu_y
\end{bmatrix}
+ \frac{t - \mu_z}{\sigma_{zz}}
\begin{bmatrix}
\sigma_{xz} \\ \sigma_{yz}
\end{bmatrix}
=
\begin{bmatrix}
\mu_u \\ \mu_v
\end{bmatrix}.
\]
This relation reflects how the mean position of the Gaussian footprint in the image plane moves as we slice through different depth levels.

The corresponding conditional covariance matrix is given by:
\[
\Sigma_{uv \mid t}
=
\begin{bmatrix}
\sigma_{xx} & \sigma_{xy} \\
\sigma_{xy} & \sigma_{yy}
\end{bmatrix}
- \frac{1}{\sigma_{zz}}
\begin{bmatrix}
\sigma_{xz} \\ \sigma_{yz}
\end{bmatrix}
\begin{bmatrix}
\sigma_{xz} & \sigma_{yz}
\end{bmatrix}.
\]
This matrix accounts for how the uncertainty in $u$ and $v$ directions reduces once the depth $t$ is fixed. The resulting conditional distribution defines a 2D elliptical Gaussian footprint on the slice at depth $t$, characterizing how the 3D Gaussian projects onto the imaging plane at that slice.

Hence, the full 3D Gaussian density can be expressed as the product:
\[
p(u, v, t) = p(t) \cdot p(u, v \mid t),
\]
which provides a factorized and computationally efficient way to model volumetric data, particularly useful in applications like slice-based rendering or tomographic reconstruction, where one often processes individual planes of a 3D volume.

\newpage
\section{Simulation on Gaussian Selection via Conditional Splat}
To further analyze the accuracy of conditional splat, we compare the true 3D Gaussian iso-probability contours on a slice with the conditional splat approximation. we generate a 3D Gaussian distribution and extract its true iso-probability contours by intersecting the Gaussian density with a fixed slice. The conditional splat approximation is then computed from the corresponding 2D ellipse splat, scaled with the marginal factor.

\begin{figure}[htbp]
    \centering
    \includegraphics[width=0.25\textwidth]{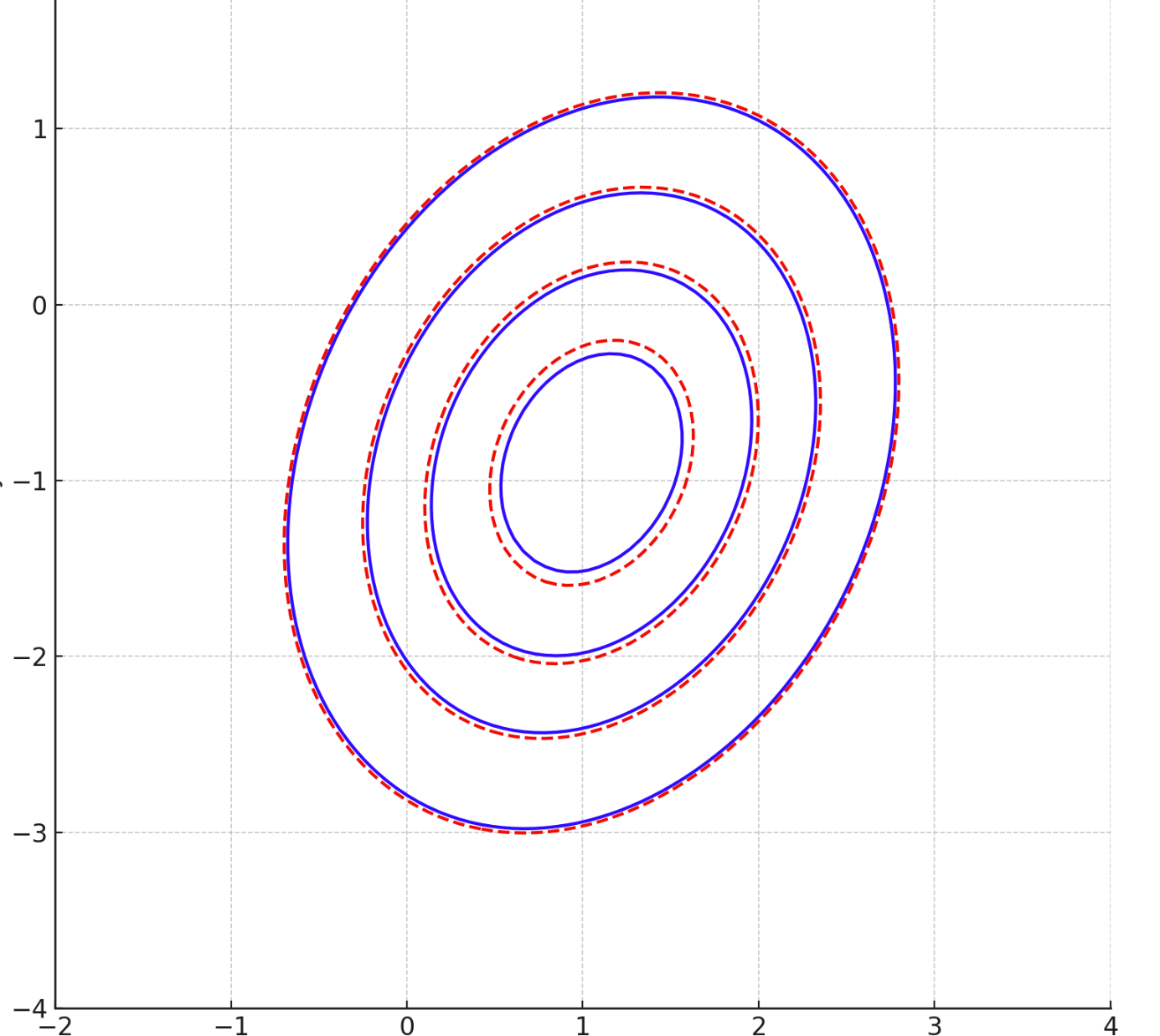}
    \caption{The conditional splat approximation (dashed red) closely matches the true 3D Gaussian iso-probability contours (solid blue).}
    \label{fig:cond}
\end{figure}

As shown in the Figure \ref{fig:cond}, the red dashed contours (conditional splat) closely overlap with the blue solid contours (true iso-probability). This demonstrates that the conditional splat provides an accurate and efficient approximation of the original 3D Gaussian's influence range.

\bibliographystyle{apalike}
\bibliography{references}	
\end{document}